
\input harvmac
%
%
%
%
\ifx\answ\bigans
\else
\output={
  \almostshipout{\leftline{\vbox{\pagebody\makefootline}}}
         \advancepageno
}
\fi
%
%
%
\def\mayer{\vbox{\baselineskip=14truept
\sl\centerline{Department of Physics 0319}%
\centerline{University of California, San Diego}
\centerline{9500 Gilman Drive}
\centerline{La Jolla, CA 92093-0319}}}
%
%

%
%
%
%
\def\abstract#1{\centerline{\bf Abstract}\nobreak
\medskip\nobreak\par #1}
%
%
%
%
%
%
%
%
%
\def\inv{^{\raise.15ex\hbox{${\scriptscriptstyle -}$}\kern-.05em 1}}
\def\lbar{{\lower.35ex\hbox{$\mathchar'26$}\mkern-10mu\lambda}}


%
%
%
%
\def\dsl{\,\raise.15ex\hbox{/}\mkern-13.5mu D}

\def\delsl{\raise.15ex\hbox{/}\kern-.57em\partial}
\def\Ksl{\hbox{/\kern-.6000em\rm K}}
\def\Asl{\hbox{/\kern-.6500em \rm A}}
\def\Dsl{\hbox{/\kern-.6000em\rm D}} 
\def\Qsl{\hbox{/\kern-.6000em\rm Q}}
\def\gradsl{\hbox{/\kern-.6500em$\nabla$}}
%
%
\def\lspace{\ifx\answ\bigans{}\else\qquad\fi}
\def\lbspace{\ifx\answ\bigans{}\else\hskip-.2in\fi} 
%
%
\def\boxeqn#1{\vcenter{\vbox{\hrule\hbox{\vrule\kern3pt\vbox{\kern3pt
        \hbox{${\displaystyle #1}$}\kern3pt}\kern3pt\vrule}\hrule}}}
%
%
\def\mbox#1#2{\vcenter{\hrule \hbox{\vrule height#2in
\kern#1in \vrule} \hrule}}
%
%
%
%

%
%
%
%
%

%

%
%

\def\darr#1{\raise1.5ex\hbox{$\leftrightarrow$}\mkern-16.5mu #1}

%
%
\def\frac#1#2{{\textstyle{#1\over #2}}} 
%
%
%
%

%
%
%
%

%
%
\def\ltap{\ \raise.3ex\hbox{$<$\kern-.75em\lower1ex\hbox{$\sim$}}\ }
\def\gtap{\ \raise.3ex\hbox{$>$\kern-.75em\lower1ex\hbox{$\sim$}}\ }
\def\gl{\ \raise.5ex\hbox{$>$}\kern-.8em\lower.5ex\hbox{$<$}\ }
\def\roughly#1{\raise.3ex\hbox{$#1$\kern-.75em\lower1ex\hbox{$\sim$}}
}
%
%

%

%

\relax

\def\dpart{\partial\kern .5ex\llap{\raise
1.7ex\hbox{$\leftrightarrow$}}\kern -.7ex {_\mu}}

\def\frac#1#2{{\textstyle{#1 \over #2}}}

\def\s2weak{\sin^2\theta_{\rm w}}

\def\({\left(}\def\){\right)}

\def\mayer{\vbox{\baselineskip=14truept\sl
\centerline{Department of Physics, 9500 Gilman Drive 0319}
\centerline{University of California, San Diego}
\centerline{La Jolla, CA 92093-0319}}}

\def\Queens{\vbox{\baselineskip=14truept
\sl\centerline{Department of Physics, Stirling Hall}
\centerline{Queen's University}
\centerline{Kingston, Canada, K7L 3N6}}}
\def\Duke{\vbox{\baselineskip=14truept
\sl\centerline{Department of Physics}
\centerline{Duke University, Durham, NC 27708-0305}}}

\def\[{\left[}
\def\]{\right]}
\def\({\left(}
\def\){\right)}

\noblackbox
 at 12truept
\vskip 1.in
\centerline{\titlefont Probing Hadronic Structure with}
\medskip
\centerline{\titlefont The Decay $\Delta\rightarrow Nl^+l^-$}
\bigskip
\centerline{Malcolm N.\ Butler}
\bigskip
\Queens
\bigskip
\centerline{Martin J.\ Savage\footnote{$^{\dagger}$}
{SSC Fellow}}
\bigskip
\mayer
\bigskip
\centerline{Roxanne P.\ Springer}
\bigskip
\Duke
\bigskip
\centerline{{\bf Abstract}}
\medskip

We compute the branching ratio for $\Delta\rightarrow Ne^+e^-$
and $\Delta\rightarrow N\mu^+\mu^-$
in chiral perturbation theory and find that both decays should be
observable at CEBAF.
With sufficiently low thresholds on the $e^+e^-$ invariant mass
a branching ratio of $\sim 10^{-5}$ may be observed for
$\Delta\rightarrow Ne^+e^-$.
For the $\Delta\rightarrow N\mu^+\mu^-$ decay mode
we predict a branching ratio of  $3\times 10^{-7}$.
The dependence of the M1 and E2 amplitudes on the momentum transfer
will provide a useful test of chiral perturbation theory which predicts
$\sim 20\%$ variation over the allowed kinematic range.

\vfill
\line{UCSD/PTH 93-06, QUSTH-93-02, Duke-TH-93-48 \hfil March1993}
\line{hep-ph/9303015\hfil}
\eject

The radiative decays of the $\Delta$ and hyperon resonances can
provide
valuable insight into the nature of baryon structure and dynamics;
a laboratory for probing nonperturbative QCD.
These decays have
recently been studied from the standpoint of heavy baryon chiral
perturbation theory  (HB$\chi$PT) \ref\buta{M. N. Butler, M.J. Savage
and R.P. Springer, Strong and Electromagnetic Decays of the Baryon
Decuplet, UCSD/PTH 92-37, QUSTH-92-05, Duke-TH-92-44 (1992).}
\ref\butb{M. N. Butler, M.J. Savage and R.P. Springer, E2/M1 Mixing
Ratio of $\Delta\rightarrow N\gamma$ and Hyperon Resonance Radiative
Decay, UCSD/PTH 93-04, QUSTH-93-01, Duke-TH-93-47 (1993).}\ and from
quenched QCD lattice gauge theory\ \ref\lein{D.B. Leinweber, T.
Draper
and R.M. Woloshyn, Baryon Octet to Decuplet Electromagnetic
Transitions, U. MD PP \#93-085, U. KY PP \#UK/92-09, TRIUMF PP
\#TRI-PP-92-120 (1992).}.  The SU(3) allowed transitions should be
easily observable at CEBAF.  The SU(3) forbidden modes, having
branching ratios substantially smaller than those of the
SU(3) allowed modes, will be
on the verge of observability.  In \butb\  the E2/M1 mixing ratio (ratio of
reduced matrix elements for electric quadrupole to magnetic dipole
radiation) was computed including final state interactions arising
from
soft pion rescattering.  The mixing ratios were found to be large,
with
substantial imaginary parts.

Further information about the radiative vertex can be gleaned from a
study of $\Delta\to Nl^+l^-$, where $l$ is an electron or muon.
The off--shell behaviour of this vertex
can be determined  by studying the decay rate of $\Delta\to Nl^+l^-$
as a function of the lepton pair invariant mass.
In this work we will extend
the calculations of ref.\ \buta\ and \butb\ to compute the rate for
this
process.  Both $\Delta\rightarrow Ne^+e^-$ and
$\Delta\rightarrow N\mu^+\mu^-$ should be observable at CEBAF;
 if sufficiently low invariant mass lepton pairs can be resolved
then the branching fraction for  $\Delta\rightarrow Ne^+e^-$
could be as large as $\sim10^{-5}$.

We are working in the limit that the baryon mass, $M_B$, is much larger
than
any momentum transfer in the process, such as the pion mass, photon
energy,
etc.\ (for a review see
\ref\jmhungary{E.  Jenkins and A. Manohar, Proceedings of the
workshop
on``Effective Field Theories of the Standard Model,'' ed.  U.
Meissner, World Scientific (1992)}).  In
such a limit the four-velocity of the baryon, $v_\mu$, becomes a good
quantum number, as does its spin.  In computing the
matrix element for $\Delta\rightarrow Nl^+l^-$ it is most convenient
to work in the frame where the lepton pair are back--to--back with
invariant mass $s$. The M1 and E2 amplitudes, $A_1(s)$ and
$A_2(s)$
respectively, have been computed at the one loop level in \buta\ and
\butb\  and are shown in
\fig\amp{The M1 and E2 amplitudes as a function of momentum
transfer $s$.
The real part of the M1 amplitude dominates the rate and has
$\sim20\%$ variation over the entire range of kinematically
allowed values for $s$.}.
The M1 amplitude is dominated by a dimension five local
counterterm
that is fit to the decay rate for $\Delta\rightarrow N\gamma$, and by
one-loop
graphs involving kaons that give characteristic contributions of the
form
$\pi M_K/\Lambda_\chi^2$.
On the other hand, the E2 amplitude is dominated by long-distance
physics
through pion loops contributing terms of the form
$\log (M_\pi/\Lambda_\chi)$.
The extension to off--shell photons is straightforward, where
now the $s$
dependence of the kaon and pion loop graphs is retained.
With these trivial modifications the results of
\buta\ and \butb\  will be used with no further discussion.

We find that the spin
averaged square of the matrix element for $\Delta\rightarrow Nl^+l^-$
is given by
\eqn\mat{\eqalign{
 {1\over 4}|{\cal M}|^2 &= {2e^2\over 3}{M^2_B\over s^2}
(\Delta m^2-s)\bigg[  |A_1(s)|^2\left( s(1+\beta_l^2\cos^2\theta) +
4m_l^2\right)   \cr
& \left. + {1\over 5} |A_2(s)|^2
\left(4{s^2\over\Delta m^2}(1-\beta_l^2\cos^2\theta)+
3s(1+\beta_l^2\cos^2\theta)
+12m_l^2\right)\right] }\ \ \ \ ,}
where $\theta$ is the angle between the lepton momentum and the
baryon
momentum, $\Delta m$ is the $\Delta-N$ mass difference, $m_l$ is the
lepton mass, and $\beta_l=\sqrt{1-{4m_l^2/ s}}$  is the lepton
velocity.  The decay rate is then given by
\eqn\width{\eqalign{\Gamma_{l^+l^-} = {\alpha\over 36\pi^2} &
\int_{4m_l^2}^{\Delta m^2} ds {1\over s^2} \beta_l
(\Delta m^2-s)^{3/2}
\bigg[  |A_1(s)|^2(s+2m_l^2)   \cr
& \left. + {1\over 5} |A_2(s)|^2\left( 2{s^2\over\Delta m^2} +
3s + 2m_l^2\left(3+{2s\over\Delta m^2}\right)\right) \right] } \ \ \
\ \ .}

We have presented $d\Gamma_{l^+l^-}/ds$ as a function of the lepton
pair invariant mass $s$ in
\fig\diff{Differential decay rates for the two processes $\Delta\to
Ne^+e^-$ and
$\Delta\to N\mu^+\mu^-$.}.
The contribution from $A_2(s)$ is much smaller than that from
$A_1(s)$, as
expected.   Unfortunately, even deviations in the M1 angular
distribution arising
from the small admixture of the E2 component are too small to be
measured
experimentally and consequently this will not be a feasible method
to independently determine the E2/M1 mixing ratio, $\delta_{\rm
E2/M1}$.
We have also presented the branching ratio for $\Delta\to Ne^+e^-$
as
a function of the minimum detectable lepton pair invariant mass,
$s_{\rm min}$, in
\fig\tot{The branching ratio for $\Delta\to Ne^+e^-$, as a   function
of the threshold invariant mass $s_{\rm min}$.}.

There is an uncertainty of approximately 20\% in these results
arising from
uncertainties in the  coupling constants used to determine $A_1(s)$
and $A_2(s)$.  This issue is discussed in some detail in refs.\
\buta\ and \butb.
Since the computation was done at the one-loop level there  are $\sim 30\%$
systematic uncertainties arising from our neglect of terms
higher order in the chiral expansion.

For $s_{\rm min}=4 m_l^2$ (the minimum value corresponding to both
leptons being produced at rest), both $\Delta\to Ne^+e^-$ and
$\Delta\to\mu^+\mu^-$ have branching ratios which will be easily
observable  at CEBAF; approximately
$5\times 10^{-5}$ and $3\times 10^{-7}$  respectively.
Obviously, $s_{\rm min}=4 m_e^2$ is unrealistically low due to
insurmountable backgrounds from soft QED processes, but an
observable branching ratio of $\sim 10^{-5}$ for  $\Delta\to Ne^+e^-$
may not be unrealistic.
We urge experimentalists planning to study the decay modes of
the $\Delta$ to push their thresholds on the lepton pair invariant mass
to as low a value as possible.
This is crucial for making an accurate determination of the $s$ dependence
of the M1 amplitude.

In conclusion, we have computed the rates for
$\Delta\rightarrow Ne^+e^-$ and $\Delta\rightarrow N\mu^+\mu^-$
in chiral perturbation theory and find that both modes should be
easily observable at CEBAF.  Measurement of the differential cross
section
for these decays will provide detailed invaluable information on the
``formfactor" of the M1 amplitude.  Chiral perturbation theory predicts
only a slight variation, $\sim 20\%$, with respect to the invariant mass of the
lepton pair over the entire allowed kinematic range arising from the kaon loops
(the local counterterm has no $s$ dependence) as shown in \amp.
Determination of this $s$ dependence
would be a nice test of chiral perturbation theory.

We also note that a similar analysis could be done for
$\Sigma^*\to\Sigma l^+l^-$ and $\Xi^*\to\Xi l^+l^-$, though they
would be suppressed relative to real photon decay by the same
amount as the $\Delta\to Nl^+l^-$ decay.
Even more interesting would be a measurement of the $s$ dependence of
the SU(3) forbidden transitions (which do not receive a contribution
from the local counterterm and consequently will have a much
stronger $s$ dependence than the SU(3) allowed transitions).
Unfortunately, it seems unlikely that CEBAF will produce a large
enough number of strange resonances to make such measurements
feasible.

\bigskip

We would like to thank F.W.\ Hersman for useful discussions.
RPS would like to thank T.J. Allen, J. Rau, and R. Tesarek
for their help.
MNB acknowledges the support of the Natural Science and
Engineering Research Council  (NSERC) of Canada.
MJS acknowledges
the support of a Superconducting Supercollider National Fellowship
from the Texas National Research Laboratory Commission under grant
FCFY9219 and was supported in part by DOE grant
DE-FG03-90ER40546.
RPS acknowledges the support of DOE grant
DE-FG05-90ER40592.

\bigskip
\bigskip
\vfil\eject

\listrefs
\listfigs
\end